\documentstyle[aaspp4]{article}

\newcommand {\etl}{\mbox{et al.}}
\newcommand {\rp} {RPr\,1}

\lefthead{A. Magazz\`u et al.}
\righthead{A brown dwarf in the Praesepe Open Cluster}

\begin{document}

\title{A brown dwarf candidate in the Praesepe Open Cluster}
\author{A. Magazz\`u}
\affil{Osservatorio Astrofisico di Catania, Citt\`a Universitaria, I-95125
Catania, Italy; antonio@ct.astro.it}
\authoremail{antonio@ct.astro.it}

\author{R. Rebolo, M.R. Zapatero Osorio, E.L. Mar{t\'\i}n\altaffilmark{1}}
\affil{Instituto de Astro{f\'\i}sica de Canarias, E-38200 La Laguna, Tenerife, 
Spain; rrl@iac.es, mosorio@iac.es, ege@iac.es}

\and

\author {S.T. Hodgkin}
\affil {Department of Physics and Astronomy, Leicester University, University 
Road, Leicester LE1~7RH, UK; sth@star.le.ac.uk}

\altaffiltext{1}{Present address: Department of Astronomy, University of 
California, Berkeley, CA 94720. USA}
\begin{abstract}
We present optical and infrared observations of \rp, a faint ($I = 21.01$) and
very red object ($I - K = 4.57$) discovered in a deep CCD survey, covering an
area of 800 square arcmin of the Praesepe open cluster. A low resolution
spectrum shows that \rp\ is a very late object, the latest object in Praesepe
for which a spectrum has been taken to date. Our estimates give a mass between
0.063 and 0.084 $M_{\sun}$, and indicate that \rp\ may turn out to be the first
brown dwarf in this cluster.

\end{abstract}


\section{Introduction}

Young, close, rich clusters make ideal hunting grounds for brown dwarfs (BDs).
The advantage of searching in these clusters is that important parameters like
age, distance, extinction, metallicity are relatively well constrained.
Moreover, the probability of detection will be increased by the higher
intrinsic luminosity of the cluster substellar objects and the higher stellar
density in the cluster with respect to the field. To date, searches for
substellar bodies have been carried out quite extensively in the Pleiades
cluster (see Hambly 1998 for a review), where recently BDs down to $0.045
M_{\sun}$ have been confirmed (Zapatero Osorio \etl\ 1997a) and many candidates
down to $0.030 M_{\sun}$ wait for confirmation.

A comparison of the lower end of the mass function of the Pleiades with that in
other open clusters would be very valuable in order to track the evolution of
very low mass objects around the substellar limit at different ages and set
constraints on predictions by models. Therefore, an extension of the search for
BDs to clusters of different ages will add to our understanding of evolution in
the substellar regime. An interesting cluster besides the Pleiades is the
Praesepe open cluster (M44). Several searches have already been performed on
this cluster. The very low main sequence of Praesepe was traced by the
photometric and proper motion survey by Hambly \etl\ (1995), and its mass
function was determined down to $0.08 M_{\sun}$ by Williams, Rieke \& Stauffer
(1995).  Recently, Pinfield \etl\ (1997) published new BD candidates, with
magnitudes down to $I = 21.5$. We discuss here low resolution spectroscopy and
IR and optical photometry of an object that may turn out to be the first bona
fide brown dwarf in Praesepe.

\section {Observations and data reduction}

The object presented in this paper was selected as a substellar candidate in a
deep $R, I$ CCD survey in the Praesepe open cluster conducted in February 1996
at the prime focus of the 2.5~m Isaac Newton Telescope, La Palma. The survey
covered an area of 800 square arcmin, complete down to $I = 21.2$, $R =
22.2$. The survey sample, data reduction and photometric analysis will be
described in a forthcoming paper. Instrumental magnitudes for \rp\ were derived
from aperture photometry, and conversion to the standard $R, I$ Cousins system
was performed observing standard stars from Landolt (1992). In agreement with
the notation in Stauffer \etl\ (1989) and Rebolo, Zapatero Osorio, \&
Mart{\'\i}n (1995), we named our object Roque Praesepe~1, hereafter referred to
as \rp, for brevity. The coordinates of \rp\ are $\alpha = 8^{\rm h} 38^{\rm m}
02 \fs 9$, $\delta = +19\arcdeg 25\arcmin 46 \farcs 7$ (Equinox 1950, Epoch
1997.11). In Figure~\ref{fin} we show a finding chart for \rp\ ($I$ image).

 Follow up infrared photometry was performed with the WHIRCAM camera at the
 4.2~m William Herschel Telescope on October 30, 1996. The camera was equipped
 with a InSb $256 \times 256$ detector array, using the $J$ and $K'$
 filters. The total integration time per filter was 200 s, each final image
 being the coaddition of 5 dithered exposures of 40~s each. Data were processed
 using standard techniques within the IRAF\footnote{IRAF is distributed by the
 National Optical Astronomy Observatory, which is operated by the Association of
 Universities for Research in Astronomy, Inc., under contract with the National
 Science Foundation.} environment. Instrumental aperture magnitudes were
 corrected for atmospheric extinction and transformed into the UKIRT system by
 observing the standard star F16 (Casali \& Hawarden 1992). The $K'$ data were
 translated into $K$ magnitudes using the conversion in Wainscoat \& Cowie
 (1992).  Such conversion makes use of the $H - K$ color. We have no $H$
 magnitude for \rp, so we adopted the average $H - K$ between Calar~3 and
 Teide~1, the reddest objects in the Zapatero Osorio, Mart{\'\i}n, \& Rebolo
 (1997b) sample, obtaining $H -K = 0.5 \pm 0.2$\@. In the Wainscoat \& Cowie
 conversion our $\pm 0.2$ error on $H -K$ translates into a $\pm 0.04$
 uncertainty, which is included in the error on $K$ quoted in Table~\ref{tfo},
 where the results of the optical and infrared photometry of \rp\ are
 summarized.

\begin{table}[h]
\caption {Photometry and spectral type of RPr 1} \label{tfo}
\begin{center}
\begin{tabular}{ll@{$\pm$}ll@{$\pm$}ll@{$\pm$}ll@{$\pm$}lc}
\tableline
~ & \multicolumn{2}{c}{$R$} & \multicolumn{2}{c}{$I$} & \multicolumn{2}{c}{$J$}
& \multicolumn{2}{c}{$K$} & Sp. Type\\
\tableline
~ & 23.5 &  0.2  &   21.01 &  0.05  & 17.7 & 0.1  & 16.44 &  0.08& $\sim M8.5$ \\
\tableline
\tableline
\end{tabular}
\end{center}
\end{table}

 Low-resolution optical spectroscopy with the red arm of the ISIS spectrograph
 was taken at the William Herschel Telescope on 1996 December 9\@. The detector
 was a TEK $1024 \times 1024$ CCD and the grating R158R gave a nominal
 dispersion of 2.9 \AA\ per pixel in the range 650-915~nm. Two spectra were
 taken (exposure times 3600 and 5300~s). Spectra were reduced by a standard IRAF
 procedure. The instrumental signature was removed making use of the standard
 G191B2B.

\section {Analysis} 

In Figure~\ref{fri}, \rp\ is located in the $K$ vs.\ $I - K$ color-magnitude
diagram, together with infrared photometry from Hodgkin \etl\ (1998) for proper
motion members from Hambly \etl\ (1995) and objects from the survey of Pinfield
at al.\ (1997). The solid lines are two isochrones from solar metallicity models
by Baraffe \etl\ (1998), for 400 and 900 Myr, shifted to the distance of
Praesepe, which we set equal to 177 pc ($m - M = 6.24$), the value determined by
Hipparcos (Mermilliod \etl\ 1997).  Although various ages have been reported for
Praesepe (cf.\ Hambly \etl\ 1995), it is generally believed (see e.g.\ Stauffer
1996) that this cluster has similar age as the Hyades ($\sim 600$~Myr). A
conservative estimate lies between 400 Myr, the value reported by Allen (1973),
and 900 Myr, as reported by Vandenberg \& Bridges (1984). Following Hambly \etl\
(1995) we will assume extinction and reddening to be zero in front of Praesepe.
We note that RPr~1 fits the photometric sequence defined by Hodgkin \etl\ (1998)
extremely well. This is very promising as far as the confirmation of cluster
membership is concerned.

In Figure~\ref{fspe} we present the best of our two low-resolution optical
spectra of \rp\@. This spectrum  (exposure time 5300~s) shows prominent TiO and VO
molecular absorption bands, typical of very late M stars, and rather strong
atomic lines of \ion{Na}{1} (818.3-819.5~nm) and \ion{K}{1} (766.5-769.9~nm),
indicating a dwarf nature. In order to derive an accurate spectral type, we have
measured the pseudocontinuum spectral ratios defined by Mart{\'\i}n, Rebolo, \&
Zapatero Osorio (1996) and the VO spectral index, defined by Kirkpatrick, Henry,
\& Simons (1995). Of the five PC indices by Mart{\'\i}n \etl\ (1996), PC5 is
outside our spectral range, and PC1, PC2 lie in the noisiest region of our
spectrum. Thus, only PC3 and PC4 have been considered. According to Mart{\'\i}n
\etl's results these two indices are the most sensitive to the spectral type. We
obtain ${\rm PC3} = 1.80$, ${\rm PC4} = 3.06$, ${\rm VO} = 1.13$, which give a
spectral type ranging between M8.2 and M9.
 
The above spread may come from the poor signal-to-noise ratio of our
spectrum. However, our measurements confirm that we are dealing with a very late
object, the latest object in Praesepe for which a spectrum has been taken to
date. In addition, the $I - K$ of \rp\ (4.57) corresponds to those of M9 dwarfs,
both in the field (Kirpatrick \& McCarthy 1994) and in the Pleiades (Zapatero
Osorio \etl\ 1997c).  We have to remark that our derivation of spectral type
could be affected by the Praesepe metallicity, slightly higher than solar (Fe/H
= 0.14, Reglero \& Fabregat 1991). Moreover, Jones \& Tsuji (1997) argue that
molecular indices should not be used for spectral classification of very
late-type objects, due to possible grain condensation in the outer layers of
their atmospheres.

We do not detect any H$\alpha$ emission in the spectrum of \rp\@. However, the
signal-to-noise ratio in the H$\alpha$ region is very low, so we estimate an
upper limit of about 3~\AA\ ($1 \sigma$) for H$\alpha$ equivalent
width. Williams \etl\ (1994) measured H$\alpha$ emission in low mass stars in
Praesepe; for M3-M4 stars (the latest type in their sample) they found
$W_\lambda ({\rm H}\alpha)$ between 2.5 and 5 \AA. To our knowledge, no data are
available for later types in this cluster. Later spectral types were
investigated by Stauffer \etl\ (1997) in the Hyades; again, equivalent widths
not exceeding 5~\AA\ were found. Our upper limit is in agreement with these
results, and is consistent --- or, at least, not in contrast --- with \rp\
membership to Praesepe. We note that in Roque~4, a Pleiades member with similar
spectral type, H$\alpha$ emission has not been observed (Zapatero Osorio \etl\
1997a). These authors point out that a membership criterion based on H$\alpha$
emission might not be valid for spectral types as late as M9.

An attempt to measure the radial velocity in the spectrum of Figure~\ref{fspe}
by cross-correlation with templates observed using the same instrumental
configuration gave a value of about zero with a $1 \sigma$ errorbar of $ \pm
40~{\rm km~ s}^{-1}$.  The radial velocity of the Praesepe cluster ($+34.3$~km
s$^{-1}$, Mermilliod \etl\ 1990) lies in the large range defined by the above
uncertainty, but obviously higher resolution spectroscopy and better
signal-to-noise ratio are required in order to confirm RPr~1 membership through
radial velocity.

From a comparison of the $K$ apparent magnitude of \rp\ with the Baraffe \etl's
(1998) isochrones, we can estimate the mass of this object. Considering a
possible range of distances between $m - M = 6.24$ (Mermillod \etl\ 1997) and
6.05 (Reglero \& Fabregat 1991) and ages between 400 and 900 Myr, we get a mass
range between 0.063 and 0.084~$M_{\sun}$\@.  Note that the main uncertainty on
the mass of \rp\ comes from the wide range of possible ages of the cluster.

An average of different temperature calibrations (Bessell \& Stringfellow 1993;
Kirkpatrick 1995; Jones \etl\ 1994; Brett 1995; Jones \etl\ 1996) gives an
effective temperature of $2400^{+250}_{-220}$~K for \rp, considering that its
spectral type is between M8.2 and M9\@. The luminosity of RPr~1 was determined
by taking an average of the values obtained from three different calibrations:
bolometric magnitude vs. $I - K$ (Tinney \etl\ 1993), $K$ bolometric correction
vs.\ $I - K$ (Tinney \etl\ 1993), and bolometric magnitude vs.\ effective
temperature (Jones \etl\ 1994). The result is $\log (L/L_\odot)= -3.60 \pm
0.16$.

\section {Discussion and conclusions}

Every point of the above analysis supports the membership of \rp\ in the
Praesepe cluster. However, we still cannot rule out that we might be dealing
with a very late M dwarf of the field. A conclusive answer would come from the
radial velocity. In absence of such measurement, further support to the \rp\
membership can be provided by the knowledge of the probability for a late M
field dwarf to appear disguised as \rp\@. Below we try to estimate how many
M8.2-M9 field dwarfs with a magnitude equal to that of \rp\ within the
photometry errors ($2 \sigma$, to be conservative) can be expected in the area
we surveyed (800 square arcmin).

From the literature (Leggett 1992; Kirkpatrick \& McCarthy 1994; Tinney
\etl\ 1995; Monet \etl\ 1992) we derive the absolute $K$ magnitudes of
M8 and M9 stars as $10.02 \pm 0.03$ and $10.41 \pm 0.13$,
respectively. Thus, we assume that any contaminating M8.2-M9 dwarf would
have $10.0 \lesssim M_K \lesssim 10.4$\@. Considering a $2 \sigma$
uncertainty around $K = 16.44$ we find that such field dwarf would lie
between 150 and 208~pc. From results in Kirkpatrick \etl\ (1994) we
estimate that the density of M8.5 field dwarfs in the solar neighborhood
is 0.0029~pc$^{-3}$. Taking into account a correction for the decrease
in density when increasing the height above the galactic plane, with a
scale height of 350 pc (Mihalas \& Binney 1981), we obtain that the
number of contaminating very late M field dwarfs that we expect to find
in our survey with the same $K$ magnitude as \rp\ is about 0.24. In
other words, we would have needed to survey an area about 4 times larger
in order to find one field dwarf with the \rp\ characteristics.

 On the other hand, we have to calculate how many BDs like \rp\ we expect in the
 cluster. In the hypothesis that the mass function of Praesepe in the substellar
 regime follows the trend found by Williams \etl\ (1995) for masses down to 0.2
 $M_{\sun}$ (a power law with slope $-1.34$), we would expect 3 objects in the
 mass range 0.063-0.084 $M_{\sun}$ in 800 square arcmin. Actually, in this
 calculation we have to use the mass range obtained adopting a unique age. If we
 use 600~Myr we obtain a mass range 0.074-0.080 $M_{\sun}$ and 1 object in 800
 square arcmin. Even in these more restrictive conditions, \rp\ has a
 probability 4 times higher to belong to Praesepe than to be a field star.

The above calculations strongly support \rp's membership. This object results
the faintest and coolest member of Praesepe and may turn out to be the first
bona fide brown dwarf in this cluster. The possible rise in the mass function of
the Pleiades beyond the substellar frontier suggested by the data of Zapatero
Osorio \etl\ (1997a,b) and our finding of \rp\ encourage a deeper extension of
the survey in a larger area of Praesepe.

\acknowledgements 

The William Herschel Telescope is operated by the Isaac Newton Group in the
Spanish Observatorio del Roque de los Muchachos of the Instituto de
Astrof{\'\i}sica de Canarias. We acknowledge support by the European Commission
through the Activity ``Access to Large-Scale Facilities'' within the programme
``Training and Mobility of Researchers'', awarded to the Instituto de
Astrof{\'\i}sica de Canarias to fund European Astronomers access to its Roque
de los Muchachos and Teide Observatories (European Northern Observatory) in the
Canary Islands. We also thank the CCI for the allocation of a fraction of the
1996 International Time at the Observatories in Canary Islands.


\clearpage

\figcaption {Finding chart of \rp\ ($I$ filter). Size is $2 \times 2$
arcmin. North is up and East at right.\label{fin}}

\figcaption{The $K$ vs.\ $I-K$ diagram for very low mass stars in Praesepe. 
Together with \rp\ (big dot with errorbars), we show two isochrones, at 400
(upper curve) and 900 Myr (lower curve), from Baraffe \etl\ (1998); data from 
Hodgkin \etl\ (1998) for proper motion members from Hambly \etl\ (1995; 
filled circles) and objects from Pinfield \etl\ (1997; open squares). In the 
isochrones, masses of 0.08 $M_{\sun}$ (stars) and 0.07 $M_{\sun}$ (crosses) 
are indicated. Errorbars are at $1 \sigma$.\label{fri}}

\figcaption{Low-resolution spectrum of \rp. Relevant atomic and molecular 
features are indicated.\label{fspe}}

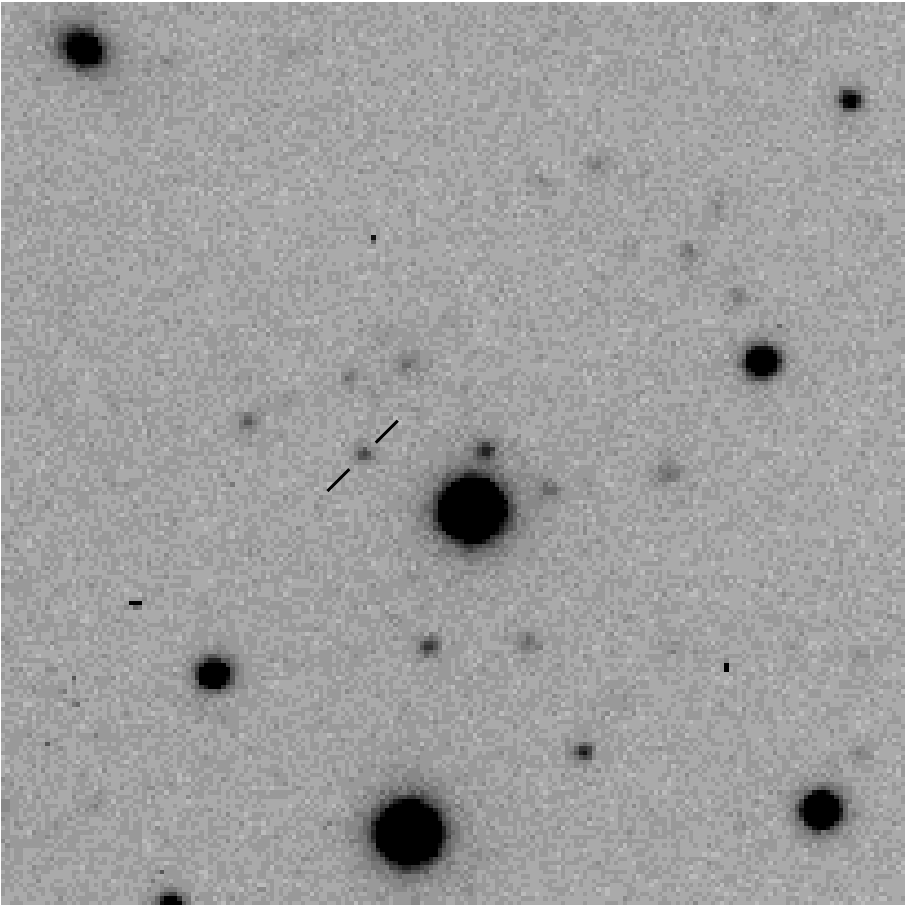
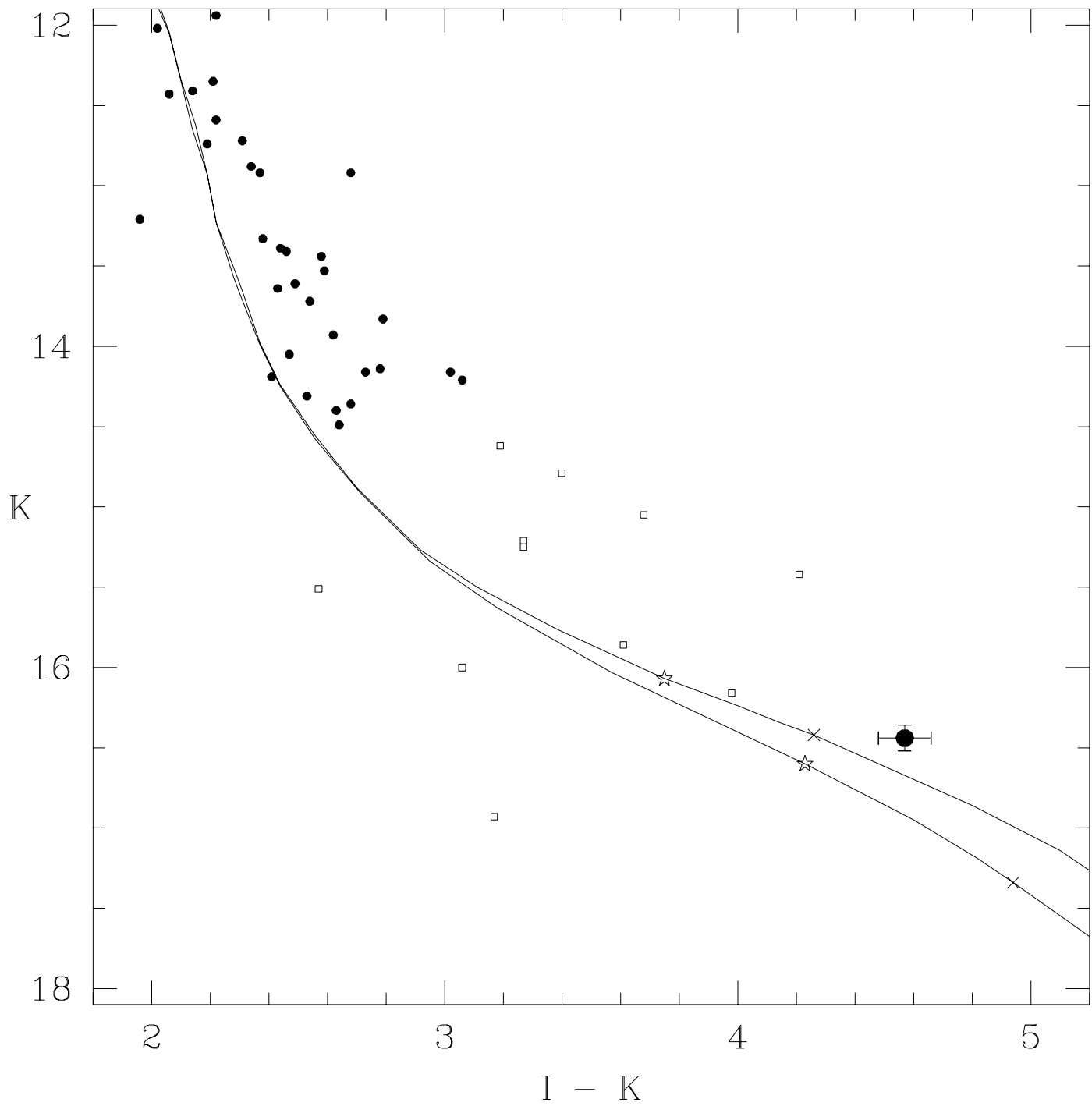
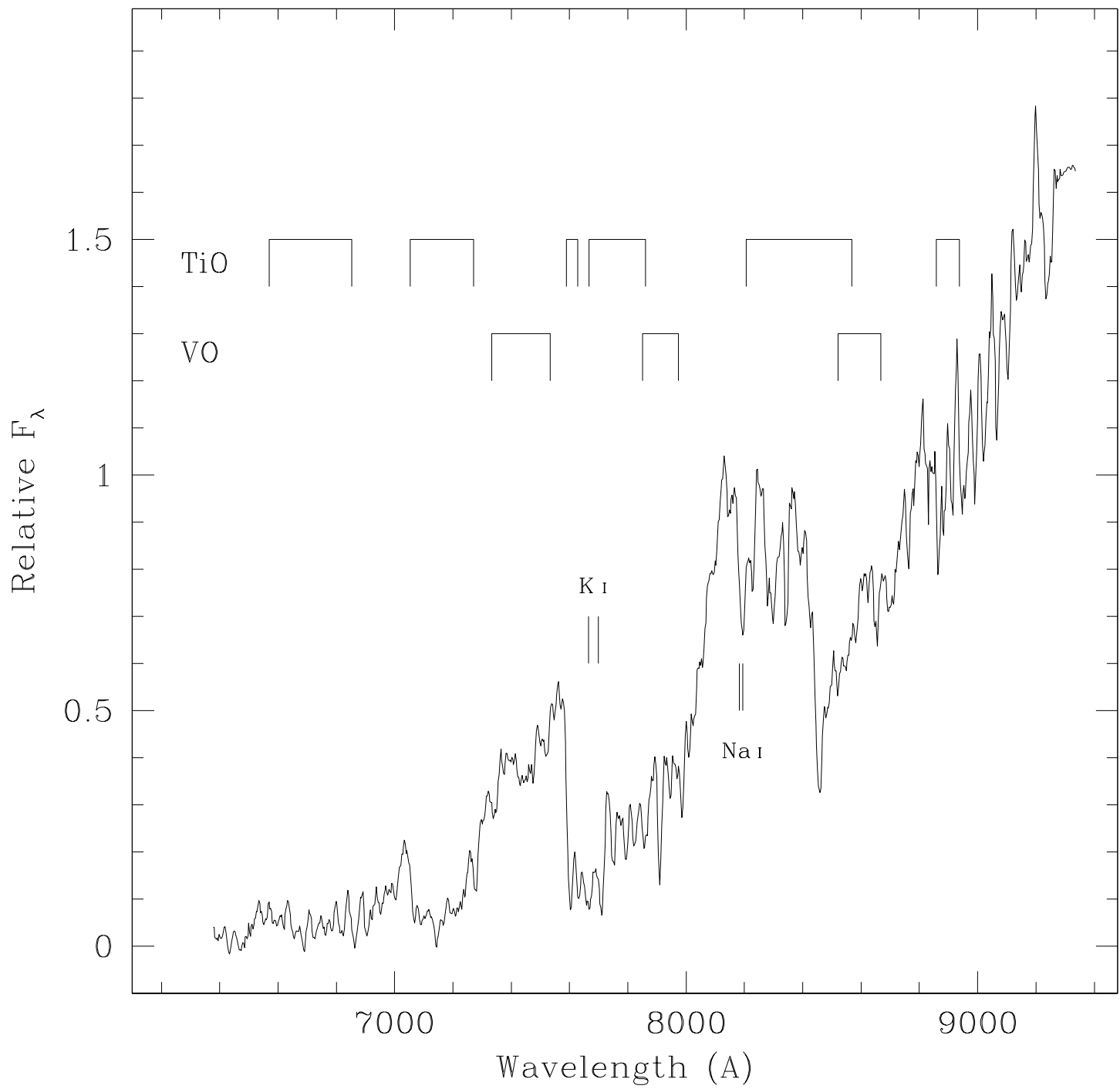
\end{document}